# Code Annealing and the Suppressing Effect of the Cyclically Lifted LDPC Code Ensemble

Chih-Chun Wang
School of Electrical & Computer Engineering
Purdue University
West Lafayette, IN 47907, USA

*Abstract*— Code annealing, a new method of designing good codes of short block length, is proposed, which is then concatenated with cyclic lifting to create finite codes of low frame error rate (FER) error floors without performance outliers. The stopping set analysis is performed on the cyclically lifted code ensemble assuming uniformly random lifting sequences, and the suppressing effect/weight of the cyclic lifting is identified for the first time, based on which the ensemble FER error floor can be analytically determined and a scaling law is derived. Both the first-order and high-order suppressing effects are discussed and quantified by different methods including the explicit expression, an algorithmic upper bound, and an algebraic lower bound.

The mismatch between the suppressing weight and the stopping distances explains the dramatic performance discrepancy among different cyclically lifted codes when the underlying base codes have degree 2 variable nodes or not. For the former case, a degree augmentation method is further introduced to mitigate this metric mismatch, and a systematic method of constructing irregular codes of low FER error floors is presented. Both regular and irregular codes of very low FER error floors are reported, for which the improvement factor ranges from $10^6$–$10^4$ when compared to the classic graph-based code ensembles.

## I. INTRODUCTION & THE MAIN RESULTS

The asymptotic analysis of low-density parity-check (LDPC) codes perfectly describes the performance when codeword length $n \geq 10^7$. Nonetheless, practical implementation generally requires short-to-moderate $n \leq 10^4$ and efficient code representation. Two major approaches of designing good finite codes consist of systematically searching through the random code ensemble or directly constructing codes under combinatorial/algebraic guidelines [1], [2]. The former approach is best represented by the progressive edge growth (PEG) construction, in which greedy/trellis-based algorithms for edge assignment have been proposed with different objective functions such as the girth [3], the extrinsic cycle degree [4], partial elimination of stopping sets [5], further criteria regarding degree 2 variable nodes [6], and the averaged error rate upper bound [7].

If one focuses on a finite code ensemble instead of a single code, recent work finds progress on different subjects including the error rate curves [8], the scaling law of the waterfall region [9], and the error-floor upper bounds [10]. As most researches along this direction, our work is based on the binary erasure channel (BEC). Existing simulations have demonstrated high correlation between the performance on BECs and on other channel models.

This paper first discusses a novel *code annealing* (CA) algorithm for finite code optimization based on the tool developed in [11] and then studies the interaction between the base code and its corresponding cyclically lifted (CL) code ensembles. A new connection between the finite code optimization and the finite-sized code ensemble analysis is established by considering the "suppressing effect" of cyclic lifting. The CA algorithm uses any existing code as a starting point and gradually enhances its performance, and thus can be seamlessly combined with any existing methods. CA is also the first efficient code optimization with the objective function being the stopping distance ($D_{\text{stp}}$), defined as the minimum size of stopping sets, instead of approximations or design heuristics. No constraint on the number of degree 2 variable nodes is assumed for CA and very good frame error rate (FER) error floor is reported for $n = 576$.

For the CL ensembles, we show that the probability that a stopping set in the base code *survives* after lifting is

$$\mathcal{O}\left(K^{-(0.5\#E - \#V + 0.5\#C_{\text{odd}})}\right),$$

where $K$ is the lifting factor, $\#V$ is the size of the base stopping set, $\#E$ is the number of edges in the subgraph induced by the base stopping set, and $\#C_{\text{odd}}$ is the number of corresponding check nodes with odd degrees. This formula prompts a definition of the *suppressing weight* $W_{\text{sup}} = 0.5\#E - \#V + 0.5\#C_{\text{odd}}$, and designing a good base code (suitable for cyclic lifting) is equivalent to maximizing the $W_{\text{sup}}$ of all small stopping sets of the base code. CA is capable of constructing a $n = 128$ regular (3,6) codes of $\min W_{\text{sup}} \geq 5$ so that all its stopping sets are experiencing at least $K^{-5}$ suppressing effects. For $K = 32$, the lifted codeword length being $n = 4096$, the suppressing factor is $2.98 \times 10^{-8}$ and results in $\text{FER}_{avg} \approx 7 \times 10^{-14}$ at erasure probability $\epsilon = 0.3$. That is, with *very high* probability, a typical code will have extremely low FER, which demonstrates a perfect match between CA and the CL ensemble. With base codes being optimized by CA, the explicitly constructed CL ensemble has similar behavior as the *expurgated* ensemble [10].

The suppressing effect analysis also sheds light on designing irregular codes of many degree 2 variable nodes, which generally has $\min W_{\text{sup}} = 0$ or 1 and does not generate satisfactory results even when codes of large stopping distances $D_{\text{stp}}$ are used as base codes. A combination of CA and

*degree augmentation* (DA) is introduced to mitigate this metric mismatch between $W_{\text{sup}}$ and $D_{\text{stp}}$, which is the first definite method of "neighborhood optimization" [12]. An $n = 576$, rate 1/2 irregular code with 62.5% degree 2 variable nodes is constructed with intact threshold performance and error floor FER $\approx 4 \times 10^{-5}$ at $\epsilon = 0.3$. This result shows that good error floors can still be obtained even in the case that there are many cycles involving only degree 2 variable nodes.

## II. Code Annealing (CA)

### A. Stopping Sets & the Asymptote of Error Floors

For a LDPC code of size $n$ and $m \times n$ parity checkmatrix $\mathbf{H}$, we denote the coded symbols as $x_1, \cdots, x_n$ and the *stopping set* as $\mathbf{s} \subseteq \{x_1, \cdots, x_n\}$ [8]. The error floor is determined by the minimum stopping sets, the size of which is defined as the stopping distance $D_{\text{stp}}$, the order of the error floor, while the total number of which is termed as $M_{\text{s}}$, the multiplicity of the error floor. $M_{\text{s}} \times \epsilon^{D_{\text{stp}}}$ then becomes an asymptote of the error floor. In [11], an efficient algorithm is provided to determine $D_{\text{stp}}$, $M_{\text{s}}$, and $\{\mathbf{s} : |\mathbf{s}| = D_{\text{stp}}\}$ for codes of $n \leq 500$.

For a code ensemble $\mathcal{C} = \{C_\alpha\}_\alpha$ where $C_\alpha$ represents the element code indexed by $\alpha$, we define the order and multiplicity of the error floor of $\mathcal{C}$ by

$$\begin{aligned} D_{\text{stp},\mathcal{C}} &= \min\{D_{\text{stp}}(C_\alpha) : \forall C_\alpha \in \mathcal{C}\} \\ M_{\text{s},\mathcal{C}} &= \mathsf{E}\{|\{\mathbf{s}(C_\alpha) : |\mathbf{s}(C_\alpha)| = D_{\text{stp},\mathcal{C}}\}|\}, \end{aligned} \quad (1)$$

where $D_{\text{stp}}(C_\alpha)$ and $\mathbf{s}(C_\alpha)$ are the stopping distance and stopping sets of $C_\alpha$. The asymptote of the ensemble FER error floor becomes $M_{\text{s},\mathcal{C}} \times \epsilon^{D_{\text{stp},\mathcal{C}}}$. One example is the classic regular (3,6) code ensemble of $n = 512$, for which we have $D_{\text{stp},\mathcal{C}} = 2$ and $M_{\text{s},\mathcal{C}} \approx 6.8 \times 10^{-2}$.

### B. The Algorithm

The CA algorithm is motivated by the following observation, and similar phenomena were observed in [13].

- For arbitrary fixed codes, the FER in the error floor region is generally dominated by only a few "bad bits," while other bits are experiencing almost error free transmission.

This typical uneven protection prompts room for improvement when properly utilized. By *locally* changing the interconnection of those "bad bits," the corresponding error patterns can be removed and the error floor performance can be drastically improved. A detailed description of CA is provided in Algorithm 1, in which the stopping set exhaustion method in [11] is adopted. This new technique is termed as *code annealing*, since most of the original code structure is kept unchanged while a small portion of edges is rearranged to generate a code of much higher performance.

The advantages of CA are three-fold. First, only local interconnection is changed during CA, and the waterfall threshold thus remains identical to its original value. Second, the local improvement nature makes CA compatible to all existing good codes or code ensembles, and can be used to construct codes from scratch or to polish existing codes. Another advantage is the use of the sufficient metric $(D_{\text{stp}}, M_{\text{s}})$ instead of indirect

**Algorithm 1** The Code Annealing (CA) Algorithm

1: Set the target order $d = 1$
2: **repeat**
3:   Construct $\mathbf{S}$, a collection of all stopping sets of size $d$.
4:   **while $\mathbf{S}$ is not empty do**
5:     Sequentially select $\mathbf{s}$, $x_a$, and $e_a$ as follows. Uniformly randomly select $\mathbf{s}$ from $\mathbf{S}$, $x_a$ from $\mathbf{s}$, and $e_a = (x_a, y_a)$ from all edges connecting $x$.
6:     Uniformly randomly select $x_b$ from var. nodes other than $\mathbf{s}$, select $e_b = (x_b, y_b)$ from all edges connecting $x_b$.
7:     Remove $e_a$ and $e_b$, and add $e'_a = (x_a, y_b)$ and $e'_b = (x_b, y_a)$ to the original bipartite graph.
8:     Find $D_{\text{stp}}$ and $M_{\text{s}}$ of the new graph.
9:     **if** $(D_{\text{stp}}, M_{\text{s}}) \succ^\dagger (d, |\mathbf{S}|)$ **then**
10:       Update $\mathbf{S}$ so that it contains all stopping sets of the new graph of size $d$.
11:     **else**
12:       Remove $e'_a$ and $e'_b$, and add back $e_a$ and $e_b$.
13:     **end if**
14:   **end while**
15:   $d \leftarrow d + 1$
16: **until** the preset computation time limit is reached.

$\dagger$ $(d_1, m_1) \succ (d_2, m_2)$ iff either $d_1 > d_2$, or $d_1 = d_2$ and $m_1 < m_2$.

ones such as the girth, the extrinsic cycle degree, etc. With this matched objective function, the benefit of CA is guaranteed and the amount of improvement is contingent only upon the efficiency of the stopping set exhaustion method. Namely, even with the random selection involved in the CA construction, it is guaranteed that there is no performance outlier in the code ensemble constructed by CA.

The edge swap trials in CA also admit parallel computing, which makes CA an appealing choice compared to other algorithms for which sequential computation is necessary.

### C. Performance Comparison — Three Examples

*Example 1:* A fixed regular (3,6) code of size 64, labelled as $C_1$, is randomly chosen from the graph-based ensemble and has $D_{\text{stp}} = 4$ and $M_{\text{s}} = 1$. After rearranging 41 out of the total 192 edges, the resulting code $C_2$ has $D_{\text{stp}} = 8$ and $M_{\text{s}} = 89$. The bit-wise error floor order is now 8 for *all* bits and *even error protection* is achieved.

Note: For any regular (3,6) code of size 64, the maximum girth is $\leq 3$ (counting only the variable nodes) while $D_{\text{stp}}$ can be improved to at least 8, as demonstrated in $C_2$. This highlights the importance of using the sufficient metric $D_{\text{stp}}$ as the objective function especially for short code optimization.

*Example 2:* $C_3$, a fixed, rate 1/2, irregular code of size 72, is randomly chosen from the graph-based ensemble with degree distributions [8], $\lambda(x) = 0.4187x + 0.1626x^2 + 0.4187x^5$ and $\rho(x) = x^5$, optimized for BECs. $C_3$ has $D_{\text{stp}} = 2$ and $M_{\text{s}} = 1$. After rearranging 61 out of the total 216 edges, the new code $C_4$ has $D_{\text{stp}} = 8$, $M_{\text{s}} = 71$. In $C_4$, all bits except one have error floor order 8 and the exception bit has order 9. The error protection is again balanced.

*Example 3:* A rate 1/2 irregular code of $n = 576$ is drawn from the same degree distributions as in Example 2. $(D_{\text{stp}}, M_{\text{s}})$ is improved from $(2, 2)$ to $(11, 58)$ after rearranging 247 out of the total 1726 edges. A significantly lower error floor is observed in Fig. 1(c) and the threshold performance is

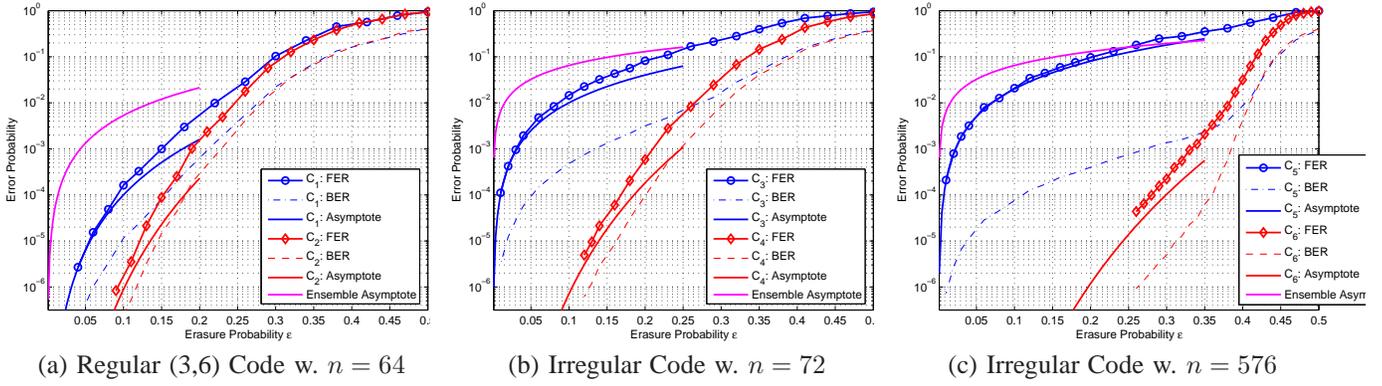

(a) Regular (3,6) Code w. $n = 64$     (b) Irregular Code w. $n = 72$     (c) Irregular Code w. $n = 576$

Fig. 1. Code performance after code annealing.

intact, which is best seen by the corresponding bit error rate (BER) curves.

As plotted in Fig. 1(a–c), the error floors of both the FER and the BER curves are significantly lowered by CA. In sum, the local edge optimization of CA is able to locate a *very good* code in the neighborhood of the starting point.

## III. THE SUPPRESSING EFFECT OF THE CL ENSEMBLE

### A. Preliminary — Cyclic Lifting

The idea of graph lifting can be traced back to [14] and is recently utilized in many coding researches [12]. The cyclically lifted (CL) code ensemble can be constructed by a base parity check code (or equivalently by a base parity check matrix $\mathbf{H}$), a lifting factor $K \in \mathbb{N}$, and randomized lifting sequences $\mathbf{l}$. To be more explicit, we first index the 1's in $\mathbf{H}$ by $1_1, 1_2, \cdots, 1_{n_e}$, where $n_e$ is the total number of 1's. Uniformly randomly select a lifting sequence $\mathbf{l} = (l_1, l_2, \cdots, l_{n_e})$ from the set $\{0, 1, \cdots, K-1\}^{n_e}$. Let $\Gamma^l$ denote a $K \times K$ permutation matrix obtained from circularly shifting an identity matrix $\mathbf{I}$ a distance $l$ to the left. The lifted code based on the lifting sequence $\mathbf{l}$ then has a parity check matrix $\mathbf{H}'$ of size $Km \times Kn$, which is obtained by replacing all 0's by a $K \times K$ zero matrix $\mathbf{0}$ and by replacing $1_i$ by $\Gamma^{l_i}$ for all $i \in \{1, \cdots, n_e\}$. It has been shown that the lifted structure admits efficient representation and encoding methods.

### B. The First Order Suppressing Effect

Let $C_L$ denote the lifted code obtained from a base code $C_B$. For any stopping set $\mathbf{s}_L$ of the lifted code $C_L$, rename all $x_i \in \mathbf{s}_L$ by $x_{\lfloor \frac{i-1}{K} \rfloor + 1}$ and use $\mathbf{s}_{L/K}$ to denote the result after this conversion. We construct $\mathbf{s}_B$ from $\mathbf{s}_{L/K}$ by removing the repeated nodes in $\mathbf{s}_{L/K}$. We then have

*Lemma 1:* $\mathbf{s}_B$ is also a stopping set of the base code $C_B$. In words, every stopping set in the lifted code corresponds to a unique stopping set in the base code. We then have the following corollary:

*Corollary 1:* The sopping distance of the lifted code is no less than that of the base code.

Based on *Lemma 1*, instead of selecting $\mathbf{s}_L$ first as described in the previous construction, we sometimes directly choose $\mathbf{s}_B$ as a stopping set of $C_B$ and define $\mathbf{s}_L$ as a stopping set of $C_L$ that is compatible with $\mathbf{s}_B$. The suppression of different orders is then defined as follows.

*Definition 1 (Suppression of Different Orders):* If $\mathbf{s}_{L/K}$ contains no repeated nodes, we define $\mathbf{s}_L$ as the survival of $\mathbf{s}_B$ against the first order suppression. Sometimes we use "$\mathbf{s}_L$ is a first order survival of $\mathbf{s}_B$" as shorthand. The high order survivals are defined as those $\mathbf{s}_L$ not a first order survival.

With a fixed lifting factor $K$ and a uniformly randomly chosen lifting sequence, we then quantify the first order suppressing effect in the following theorem.

*Theorem 1:* For any $\mathbf{s}_B$, let $\mathbf{S}_{L,1}$ denote the collection of all first order survivals of $\mathbf{s}_B$. We then have

$$\mathsf{E}\{|\mathbf{S}_{L,1}|\} = K^{\#V - \#E} \prod_{j=1}^{\#C} \left( \sum_{t=0}^{\min(K, \deg(c_j))} (-1)^t \binom{\deg(c_j)}{t} \binom{K}{t} t! (K-t)^{\deg(c_j)-t} \right),$$

where $\#V = |\mathbf{s}_B|$, $\#E$ and $\#C$ are the numbers of check nodes and edges in the subgraph (of $C_B$) induced by $\mathbf{s}_B$, and $c_j$ and $\deg(c_j)$ represent the individual check node and its degree in the induced subgraph. For future reference, we define $\mathsf{E}\{|\mathbf{S}_{L,1}|\}$ as the first order suppressing effect of $\mathbf{s}_B$.

*Proposition 1:* For any $\mathbf{s}_B$, $\mathsf{E}\{|\mathbf{S}_{L,1}|\}$ is of the order

$$K^{-(0.5 \#E - \#V + 0.5 \#C_{\text{odd}})},$$

where the new notation $\#C_{\text{odd}}$ is the number of check nodes of odd degrees in the subgraph induced by $\mathbf{s}_B$.

*Theorem 1* and *Proposition 1* are derived from the enumerator polynomials for the CL ensemble.

Define the suppressing weight of $\mathbf{s}_B$ as $W_{\text{sup}} := 0.5 \#E - \#V + 0.5 \#C_{\text{odd}}$. From the above results, the larger $W_{\text{sup}}$ is, the smaller expected number of the first order survivals will be after cyclic lifting. Furthermore, when combined with (1), one can obtain the following error floor scaling law.

*Theorem 2:* For the cyclically lifted ensemble $\mathcal{C}_{\text{CL}}$, we have

$$D_{\text{stp}, \mathcal{C}_{\text{CL}}} = D_{\text{stp}}(C_B)$$

$$M_{\text{s}, \mathcal{C}_{\text{CL}}} = K^{-\min\{W_{\text{sup}}: \text{ all min. } \mathbf{s}_B\}} \cdot M_{\text{s}}(C_B)(1 + o(1)).$$

For base codes of minimum variable node degree $\geq 3$, $W_{\text{sup}} \geq 0.5 D_{\text{stp}}$ for all $\mathbf{s}_B$. Using a $n = 128$, $(D_{\text{stp}}, M_{\text{s}}) = (10, 40)$, regular (3,6) code obtained from CA, the ensemble averaged FER error floor can be significantly lower even for small lifting factors $K = 4$–$8$ ($Kn = 512$–$1024$). The comparison of ten $n = 512$ sample codes chosen from the CL and the classic ensembles is provided in Fig. 2. With a much lower averaged FER error floor, $(D_{\text{stp},\mathcal{C}_{\text{CL}}}, M_{\text{s},\mathcal{C}_{\text{CL}}}) = (10, 0.015)$, situated at $8.74 \times 10^{-8}$ when $\epsilon = 0.3$, the performance of the CL ensemble is highly concentrated when FER $\geq 10^{-6}$, and no error floor can be observed by the Monte-Carlo (MC) simulation. Without those apparent performance outliers as in the classic ensemble, the CL ensemble behaves very much like the *expurgated* ensemble. For comparison, under MC simulations, all 10 randomly sampled CL codes are of performance indistinguishable from the best existing $n = 504$ regular codes optimized by PEG [3].

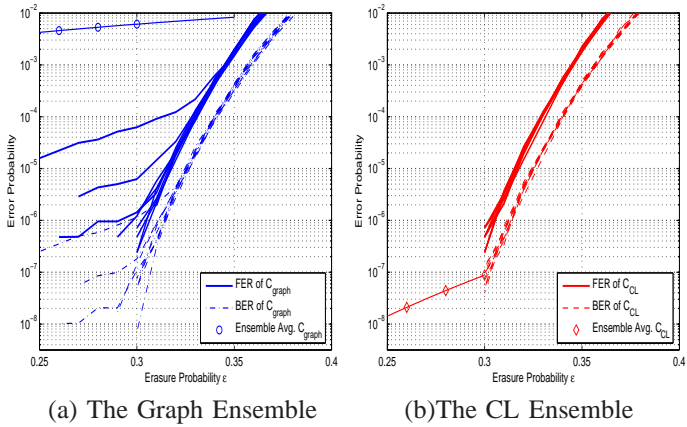

Fig. 2. Comparison between the classic graph-based ensemble and the CL ensemble with $K = 4$, base code $(D_{\text{stp}}, M_{\text{s}}) = (10, 40)$. The codeword length for both ensembles is $n = 512$, and 10 codes are randomly chosen from each ensemble.

From our experiments, almost all $\mathbf{s}_L$ of sizes close to $D_{\text{stp}}(C_L)$ are first order survivals, which can be explained by the following two reasons. By definition, high-order survivals have repeated nodes in $\mathbf{s}_{L/K}$. Therefore, for any $\mathbf{s}_B$, the sizes of high order survivals are strictly larger than the size of the first order survivals if there is any, and $D_{\text{stp}}(C_L)$ is more likely to be determined by the first order survivals. Secondly, high order suppressing effects are decided not only by combinatorial arguments as in *Theorem 1* but also by algebraic constraints as in *Theorems 3 and 4*, which further limits the chances of seeing high-order survivals of small sizes.

### C. The High Order Suppressing Effect

The expression to quantize the high order suppressing effect is generally more involved. Let $\#V_L$, $\#C_L$, and $\#E_L$ represent the number of variable nodes, check nodes, and edges in the subgraph of $C_L$ induced by $\mathbf{s}_L$, and define $\#V_B$, $\#C_B$, and $\#E_B$ similarly. We then have

*Theorem 3 (Algebraic Lower Bound):* For any $\mathbf{s}_B$, let $\mathbf{S}_{L,\geq 1}$ be the collection of $\mathbf{s}_L$ sharing the same $\#V_L$, $\#C_L$, and $\#E_L$ values. It is either $\mathbf{S}_{L,\geq 1}$ is empty or

$$\mathsf{E}\{|\mathcal{S}_{L,\geq 1}|\} \geq \text{const} \cdot \max(K^{-(\#E_L - \#V_L - \#C_L)}, K^{-(\#E_B - \#V_B - \#C_B)})$$

for some const not depending on $K$.

$\mathbf{S}_{L,\geq 1}$ contains either first order survivals or high order survivals exclusively, depending on whether $\#V_L = \#V_B$ or not. *Theorem 3* is derived from the linear homogeneous equations associated with stopping set configurations such that the high order survivals correspond to the number of non-zero solutions. In many occasions when the number of equations is larger than the number of free variables, $\mathbf{S}_{L,\geq 1}$ is empty, which significantly limits the high order suppressing effect. *Theorem 3* is tight when considering the first order suppressing effect for which $\#V_L = \#V_B$, $\#E_L = \#E_B$, and $\#C_L \leq 0.5\#E_B - 0.5\#C_{B,\text{odd}}$.

For any $\mathbf{s}_B$, the expected number of high order survivals can be determined by directly solving the rank of the linear equations imposed by $\mathbf{s}_B$. This computation however has to be repeated when different $\mathbf{s}_B$ are considered. A simpler upper bound will be provided, which is applicable to all possible choices of $\mathbf{s}_B$.

Let the repetition $R(x_i)$ of any $x_i \in \mathbf{s}_B$ denote the number of repeated appearances of $x_i$ in $\mathbf{s}_{L/K}$. Let $\mathbf{x}_{\text{o.d.}} \subseteq \mathbf{s}_B$ denote a subset of variable nodes such that by revealing the values of all elements in $\mathbf{x}_{\text{o.d.}}$, the rest variable nodes $\mathbf{s}_B \backslash \mathbf{x}_{\text{o.d.}}$ can be successfully decoded in a non-increasing order of the repetition. Namely, if $x_i$ is the next to be decoded variable node via check node $y_j$, there exists at least another variable node $x'_i \neq x_i$ connected to $y_j$ such that $R(x'_i) \geq R(x_i)$. The subscript "o.d." stands for ordered decoding. We then have the following theorem.

*Theorem 4 (An Algorithmic Upper Bound):* If we slightly abuse the notation and use $\mathbf{S}_{L,\geq 1}$ to denote the collection of $\mathbf{s}_L$ compatible to a given repetition pattern $\{R(x)\}_{x \in \mathbf{s}_B}$, there exists a const not depending on $K$ such that $\forall K, \forall \mathbf{x}_{\text{o.d.}}$,

$$\mathsf{E}\{|\mathbf{S}_{L,\geq 1}|\} \leq \text{const} \cdot K^{\left(\sum_{x \in \mathbf{x}_{\text{o.d.}}}(R(x)-1)\right)} K^{-W_{\sup}}.$$

Similarly, $\mathbf{S}_{L,\geq 1}$ contains either first order survivals or high order survivals exclusively, depending on whether $\{R(x)\}$ are all 1's. *Theorem 4* is tight both for the first order survivals and for the parallel high order survivals with $|\mathbf{x}_{\text{o.d.}}| = 1$ and $R(x) = R, \forall x \in \mathbf{s}_B$. *Theorem 4* further confirms the role of $W_{\sup}$ for high-order suppression.

## IV. IRREGULAR CODE OPTIMIZATION

For a base code of high $D_{\text{stp}}$ but many degree 2 nodes, the lifted code does not generate satisfactory results since there are generally many $\mathbf{s}_B$ with $W_{\sup} = 0$ or $0.5$, which benefit little from lifting. A degree augmentation (DA) method is provided to mitigate this mismatch between $D_{\text{stp}}$ and $W_{\sup}$.

- Base code optimization (DA+CA): Arbitrarily choose $d_u \in \mathbb{N}$. For those variable nodes of $\deg(x) \geq 3$, construct $d_u(\deg(x)-2)$ pairs of auxiliary variable/check nodes $(x_a, y_a)$ such that $y_a$ is connected to $x$ and $x_a$, $\forall a \in [1, d_u(\deg(x)-2)]$. Perform CA on the new graph. After CA, remove those auxiliary $(x_a, y_a)$.
- Perform cyclic lifting
- Perform CA on the lifted code, focusing on the "local lifting sequence" instead of the "local edge configuration."

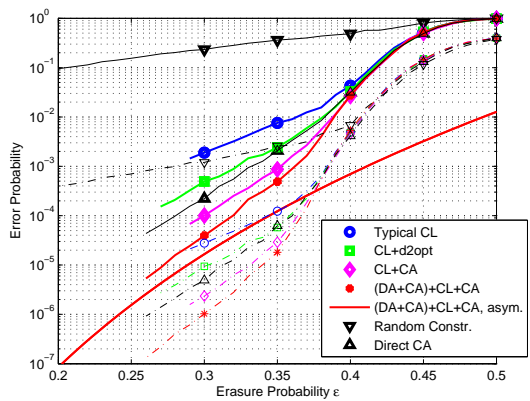

Fig. 3. Comparison among different CL constructions. "Typical CL" is randomly chosen from the CL ensemble using $C_4$ as the base code. "CL+d2opt" is obtained by applying girth optimization over degree 2 nodes. By further applying CA, we have "CL+CA" which achieves $D_{\text{stp}} = 12$. "(DA+CA)+CL+CA" achieves $D_{\text{stp}} = 13$ and is obtained by optimizing the base code by (DA+CA). "Random Cnstr" ($C_5$) and "Direct CA" ($C_6$) are plotted for comparison. The thicker and dashed lines are FER and BER curves respectively.

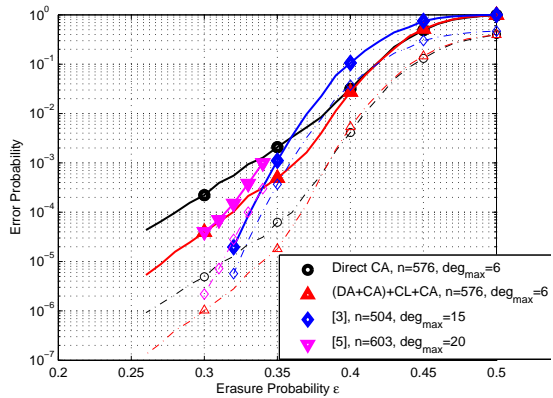

Fig. 4. Comparison between our codes in Fig. 3 and the best existing codes in [3], [5]. The thicker and dashed lines are FER and BER curves respectively.

Let $D'_{\text{stp}}$ denote the stopping distance after DA+CA but before removing the auxiliary node pairs. For sufficiently large $d_u$, the minimum $W_{\text{sup}}$ of the original graph grows linearly with respect to $D'_{\text{stp}}$. The new base code thus has high $W_{\text{sup}}$ and is suitable for cyclically lifting. The results of this DA-based method are plotted in Fig. 3. In Fig. 4, we compare the (DA+CA)+CL+CA code with some of the best existing randomized constructions in [3], [5], of which the asymptotic BEC thresholds are 0.07 and 0.22 worse than ours since their degree distributions are optimized for Gaussian channels instead. With similar error floors until $10^{-4}$ and $10^{-5}$ respectively, our code has better threshold performance even after proper calibration. This discrepancy can be explained by the local vs. global edge rearrangement, and the latter of which has the potential of changing the threshold behavior. Both their degree distributions have 47-49% of variable nodes of degree 2 while ours admits 62.5%. The large amount of degree 2 bits of our code leads to better threshold performance while limiting the lowest error floor one might achieve. In sum, our approach yields codes of low error floors comparable to the best existing codes without sacrificing the threshold performance even for codes of many degree 2 nodes. The CA applied in the last stage again eliminates any possible existence of performance outliers, a feature absent in the existing randomized constructions. Due to their distinct natures, numerical comparisons between the randomized CL construction and the algebraically constructed codes, e.g. quasi-cyclic codes [15], are deferred to a journal version of this paper.

## V. FURTHER RESEARCH DIRECTIONS

Similar suppressing effect can be observed when considering "trapping sets" for non-erasure channels. Exact formulas of the suppressing effect for the trapping sets and the corresponding CA algorithm are currently under investigation, which utilizes the trapping set exhaustion algorithm in [16].

## REFERENCES

[1] Y. Kou, S. Lin, and M. P. C. Fossorier, "Low-density parity-check codes based on finite geometries: A rediscovery and new results," *IEEE Trans. Inform. Theory*, vol. 47, no. 7, pp. 2711–2736, November 2001.

[2] B. Vasic and O. Milenkovic, "Combinatorial constructions of low-density parity-check codes for iterative decoding," *IEEE Trans. Inform. Theory*, vol. 50, no. 6, pp. 1156–1176, June 2004.

[3] X. Hu, E. Eleftheriou, and D. M. Arnold, "Regular and irregular progressive edge-growth Tanner graphs," *IEEE Trans. Inform. Theory*, vol. 51, no. 1, pp. 386–398, January 2005.

[4] T. Tian, C. Jones, J. Villasenor, and R. Wesel, "Selective avoidance of cycles in irregular LDPC code construction," *IEEE Trans. Commun.*, vol. 52, no. 8, pp. 1242–1247, August 2004.

[5] A. Ramamoorthy and R. Wesel, "Construction of short block length irregular low-density parity-check codes," in *Proc. ICC*, 2004.

[6] L. Dinoi, F. Sottile, and S. Benedetto, "Design of variable-rate irregular LDPC codes with low error floors," in *Proc. IEEE Int'l. Conf. Commun.*, no. 1. Seoul, Korea, 2005, pp. 647–651.

[7] E. Sharon and S. Litsyn, "A method for constructing LDPC codes with low error floor," in *Proc. IEEE Int'l. Symp. Inform. Theory*, July 2006, pp. 2569–2573.

[8] C. Di, D. Proietti, E. Telatar, T. J. Richardson, and R. L. Urbanke, "Finite-length analysis of low-density parity-check codes on the binary erasure channel," *IEEE Trans. Inform. Theory*, vol. 48, no. 6, pp. 1570–1579, June 2002.

[9] A. Amraoui, R. Urbanke, A. Montanari, and T. Richardson, "Further results on finite-length scaling for iteratively decoded LDPC ensembles," in *Proc. IEEE Int'l. Symp. Inform. Theory*. Chicago, 2004.

[10] T. Richardson, M. A. Shokrollahi, and R. L. Urbanke, "Finite-length analysis of various low-density parity-check ensembles for the binary erasure channel," in *Proc. IEEE Int'l. Symp. Inform. Theory*. Lausanne, Switzerland, 2002, p. 1.

[11] C. C. Wang, S. R. Kulkarni, and H. V. Poor, "Upper bounding the performance of arbitrary finite LDPC codes on binary erasure channels," in *Proc. Int'l. Symp. Inform. Theory*. Seattle, Washington, USA, July 2006.

[12] T. Richardson, "Error floors of LDPC codes," in *Proc. 41st Annual Allerton Conf. on Comm., Contr., and Computing*. Monticello, IL, USA, 2003.

[13] S. H. Lee, W. H. Lee, J. J. Bae, S. I. Lee, and E. K. Joo, "Bit probability transition characteristics of LDPC code," in *Proc. 10th International Conference on Telecommunications*. Tahiti, French Polynesia, Feb. 2003, pp. 553–557.

[14] J. Gross, "Voltage graphs," *Discrete Math.*, vol. 9, no. 3, pp. 239–246, Sept. 1974.

[15] L. Chen, J. Xu, I. Djurdjevic, and S. Lin, "Near-Shannon-limit quasi-cyclic low-density parity-check codes," *IEEE Trans. Commun.*, vol. 52, no. 7, pp. 1038–1042, July 2004.

[16] C. C. Wang, S. R. Kulkarni, and H. V. Poor, "Exhausting error-prone patterns in LDPC codes," preprint.